\documentclass[fleqn,10pt]{wlscirep}

\usepackage{amsfonts}
\usepackage{amssymb}
\usepackage{amsmath}
\usepackage{graphicx}
\usepackage{color}
\usepackage{lineno}
\usepackage{bm}

\usepackage{array}
\usepackage{makecell}

\usepackage{multirow}

\title{Specific loss power of magnetic nanoparticles (fluid) hyperthermia in non-adiabatic conditions}
\author[1,$\dagger$]{C.~A.~M.~Iglesias} 
\author[1]{J.~C.~R.~de~Ara\'{u}jo} 
\author[1]{J.~Xavier} 
\author[1]{R.~B.~da~Silva} 
\author[2]{J.~M.~Soares} 
\author[3,4]{E.~L.~Brito} 
\author[4,5]{L.~Streck} 
\author[4]{J.~L.~C.~Fonseca} 
\author[6]{C.~C.~Pl\'{a}~Cid}
\author[1]{M.~Gamino} 
\author[1]{E.~F.~Silva} 
\author[1]{C.~Chesman} 
\author[1]{M.~A.~Correa} 
\author[1]{S.~N.~de~Medeiros} 
\author[1,$\star$]{F.~Bohn} 

\affil[1]{Departamento de F\'{i}sica, Universidade Federal do Rio Grande do Norte, 59078-900 Natal, RN, Brazil}
\affil[2]{Departamento de F\'{i}sica, Universidade do Estado do Rio Grande do Norte, 59610-090 Mossor\'{o}, RN, Brazil}
\affil[3]{POLYMAT and Departamento de Química Aplicada, Facultad de Ciencias Químicas, University of the Basque Country UPV/EHU, Joxe Mari Korta Zentroa, Tolosa Hiribidea 72, 20018 Donostia-San Sebastián, Spain}
\affil[4]{Instituto de Qu\'{i}mica, Universidade Federal do Rio Grande do Norte, 59078-970 Natal, RN, Brazil}
\affil[5]{Curso de Farmácia, Faculdade Maurício de Nassau, 59080-400 Natal, RN, Brazil}
\affil[6]{Departamento de F\'{i}sica, Universidade Federal de Santa Catarina, 88040-900 Florian\'{o}polis, SC, Brazil}
\affil[$\dagger$]{iglesias@fisica.ufrn.br}
\affil[$\star$]{felipebohn@fisica.ufrn.br}


\begin{abstract}
We investigate the magnetic nanoparticles (fluid) hyperthermia in non-adiabatic conditions through the calorimetric method. 
Specifically, we propose a theoretical approach to magnetic hyperthermia from a thermodynamic point of view. 
To test the robustness of the approach, we perform hyperthermia experiments and analyze the thermal behavior of magnetite and magnesium ferrite magnetic nanoparticles dispersed in water submitted to an alternating magnetic field. 
From our findings, besides estimating the specific loss power value from a non-adiabatic process, thus enhancing the accuracy in the determination of this quantity, we provide physical meaning to parameters found in literature that still remained not fully understood, and bring to light how they can be obtained experimentally. 
\end{abstract}

\begin{document}

\flushbottom
\maketitle

\thispagestyle{empty}

\section{Introduction}
\label{Introduction}
Magnetic hyperthermia (MHT) corresponds to the effect that exploits the heat generated by magnetic nanoparticles (MNPs) when submitted to an alternating magnetic field (AMF). 
In recent decades, magnetic nanoparticles (or ferrofluid) hyperthermia has received increasing interest due to the possibility of its application as a thermal therapy in clinical trials for the treatment of cancers and other diseases, as well as in the process of thermal activated drug delivery under AMF~\cite{DelavariH.2013, Perigo2015, Branquinho2013, Abenojar2016}. 
Within this context, specific features of the magnetic nanoparticles dispersed in a carrier liquid are important, being explored to perform the drug delivery and/or destroy ill cells by heating. 

The magnetic hyperthermia effect has been extensively investigated both theoretically and experimentally. 
In magnetic nanoparticles (or ferrofluid) hyperthermia, the potential of a given magnetic material is in general evaluated through the specific loss power ($SLP$), often also denoted by the so-called specific absorption rate ($SAR$)~\cite{Garaio2015}. 
$SLP$ is simply the power generated per unit mass of the magnetic material~\cite{Carriao2017}. 
This quantity, described in terms of a linear response theory~\cite{Rosensweig2002}, is notably a function of both, material/sample properties (for instance, saturation magnetization, magnetic susceptibility, magnetic anisotropy, magnetization relaxation times, particle size, shape of the nanoparticle, particle concentration, volume and liquid viscosity) and experiment conditions (such as waveform, frequency and amplitude of the alternating magnetic field). 
For this reason, hyperthermia also appears as an important tool to provide insights on the magnetic behavior of magnetic nanoparticles, contributing specifically to the understanding of the fundamental physics associated to the magnetization dynamics in such systems with reduced dimensions. 

In a typical hyperthermia experiment, the evolution of the temperature of the nanoparticles or the ferrofluid with time is probed. 
From this measurement, $SLP$ is commonly quantified by standard calorimetric methods in which a quasi-adiabatic regime is assumed, i.e.\ the nanoparticles or the ferrofluid behave as a quasi-adiabatic system whose energy is absorbed by the magnetic material at a constant rate~\cite{Batista2015, Perigo2015, Zhang2007, Bekovic2010, Garaio2014, Abenojar2016, Bordelon2011, Carriao2017}. 
Within this picture, just the slope of the temperature curve during a short time interval after applying AMF is analyzed. 
Hence, this procedure brings intrinsic uncertainties, as well as it frequently underestimates the $SLP$ value for a suspension of MNPs~\cite{Garaio2014}. 
However, while the quasi-adiabatic regime has been widely investigated, the same effort was not intended to the analysis of non-adiabatic calorimetric methods to this end~\cite{Perigo2013, Teran2012}. 
As a consequence, many questions on the magnetic heating power of MNPs and the determination of $SLP$ are still open. 
Among them, perhaps the most remarkable doubt on the issue resides in the influence that the energy losses to the environment may have on the magnetic nanoparticles (or ferrofluid) hyperthermia response. 
In particular, the answer for this issue directly impacts the technological fields of engineering and biomedicine, given that in applications a suspension of MNPs is commonly not insulated. 
In this sense, we understand that a theoretical approach that considers parameters related to the interaction of the magnetic fluid with the environment becomes needed, thus providing further accurate estimates of $SLP$ for suspensions of magnetic nanoparticles under AMF in non-adiabatic conditions. 

In this article, we investigate the magnetic hyperthermia in suspensions of MNPs.
Specifically, we propose a theoretical approach to magnetic hyperthermia from a thermodynamic point of view. 
The model allows us to obtain the $SLP$ value from a non-adiabatic process, thus enhancing the accuracy in the determination of this quantity. 
To test the robustness of the approach, we perform hyperthermia experiments and analyze the thermal behavior of magnetite and magnesium ferrite MNPs dispersed in water submitted to an AMF. 
From our findings, besides estimating the specific loss power value, we provide physical meaning to parameters found in literature that still remained not fully understood, and bring to light how they can be obtained experimentally. 

\section{Results}
\label{Results}

\noindent{\bf Theoretical approach.} 
Here, to investigate the specific loss power of magnetic nanoparticles (fluid) hyperthermia, we focus on the temperature response of MNPs dispersed in a fluid submitted to an AMF. 
To this end, we employ a theoretical approach based basically on thermodynamics concepts, and, therefore, without the need of a microscopic description of the system. 

\noindent{\bf Mimicking an adiabatic system.} 
We start our approach to magnetic hyperthermia by presenting the well-known adiabatic model~\cite{Batista2015, Perigo2015, Zhang2007, Bekovic2010, Garaio2014, Abenojar2016, Bordelon2011, Carriao2017}, with its assumptions and limitations. 

In MHT, the heating effect of a magnetic fluid is a result of absorbing energy from the AMF and converting it into a raise of the internal energy and/or heat by eddy current losses~\cite{Bekovic2010}, hysteresis losses~\cite{Hergt2008} and relaxation losses~\cite{Bekovic2010, Brown1963, Coffey2012}. 
Generally, magnetic fluids exhibit low electrical conductivity, in a sense that the inductive heating does not arise and can in principle be neglected. 
Next, hysteresis losses are attributed to ferromagnetic/ferrimagnetic features of the particles, and they are directly related to the magnetization reversion during the magnetization process in such magnetic materials. 
At last, relaxation losses are ascribed to superparamagnetic/ferrimagnetic compounds~\cite{Perigo2013}. 
For this latter kind of loss, it is worth remarking that there are two distinct mechanisms by which the magnetization of magnetic fluids may relax after the magnetic field is removed. 
The first one is associated to the so-called Brown relaxation~\cite{Brown1963,Coffey2012}. 
In this case, the particle moves freely within the suspension, and the relaxation takes place due to the reorientation of the whole particle, being a result of the viscous friction between the rotating particle and surrounding medium~\cite{Bekovic2010}. 
The second relaxation mechanism in turn is connected to the N\'{e}el relaxation~\cite{Brown1963, Coffey2012}. 
Specifically, it consists in the reversion of the magnetic moment within the particle, once the magnetic moment overcomes an energy barrier due to the uniaxial anisotropy. 

Despite the diversity in essence, the losses in all cases come from the irreversible work undergone by the suspension due to interaction effects of the magnetic particles with the AMF. 
In order to quantify the variation of the internal energy of the suspension due to the irreversible work done by the magnetic field, we take into account the general Principle of Energy Conservation --- In an energetically isolated system, the total energy remains constant during any change which may occur in it. 
%It may be stated as follows: ``In an energetically isolated system, the total energy remains constant during any change which may occur in it.'' 
When adapted for thermodynamic processes, it is expressed by the First Law of Thermodynamics, given by 
\begin{eqnarray}
	\Delta U_{susp} = W - Q_{susp},
%	\Delta U_{susp} = W_{mag} + W_{mec} - Q_{susp} ,
	\label{1-lei}	
\end{eqnarray}
where in our context $\Delta U_{susp}$ is the variation of the internal energy of the suspension, $W$ is the irreversible work {\it undergone by the suspension}, and $Q_{susp}$ is the heat {\it lost by the suspension}. 
Here, we may split the work $W$ into two components; the first, depicted by $W_{mag}$, corresponds to the work {\it undergone by the suspension due to the interaction of the magnetic nanoparticles with the alternating magnetic field}; the second one, $W_{mec}$ is the mechanical work {\it done on the suspension}. 

For an adiabatic ($Q_{susp} = 0$) and isochoric ($W_{mec} = 0$) process, $\Delta U_{susp}$ may be written simply as 
\begin{eqnarray}
	\Delta U_{susp} = W_{mag} = C_{susp} \Delta T ,
	\label{conserv-energia}
\end{eqnarray}
where $\Delta T$ is the temperature variation of the system, i.e.\ the suspension, and $C_{susp}$ is the heat capacity of the suspension, which can be expressed in a generalized form as
\begin{eqnarray}
	C_{susp} = \sum_{j}^N m_{j} c_{j} ,
	\label{cap-term}
\end{eqnarray}
in which $m_{j}$ and $c_{j}$ are the mass and specific heat of the $j-th$ constituent (magnetic nanoparticles and fluid) of the suspension, respectively, and $N$ is the total number of constituents in the suspension. 

The specific loss power, as aforementioned, is defined as the power generated ($W/\Delta t$), where $\Delta t$ is a time interval, per unit mass of the magnetic material ($m_{np}$). 
Hence, considering Eq.~(\ref{conserv-energia}), for an adiabatic ($Q_{susp}~=~0$) and isochoric ($W_{mec}~=~0$) process, $SLP$ may be expressed as 
\begin{eqnarray}
	SLP = \dfrac{1}{m_{np}} \dfrac{W_{mag}}{\Delta t} =\frac{\left(\Delta U_{susp}/\Delta t\right)}{ m_{np}}= \dfrac{1}{m_{np}} C_{susp} \dfrac{\Delta T}{\Delta t} .
\label{slp-adiab}
\end{eqnarray}

Remarkably, Eq.~(\ref{slp-adiab}) has been addressed and employed in numerous works found in literature~\cite{Batista2015, Perigo2015, Zhang2007, Bekovic2010, Garaio2014, Abenojar2016, Bordelon2011, Carriao2017}. 
However, it is worth pointing out that this first approach to estimate $SLP$ has validity only in the quasi-adiabatic regime, i.e.\ when the system is insulated and its temperature is considered varying as a linear function with time. 
This assumption is a key factor that may affect the results, in a sense we should look with care at the $SLP$ findings. 
In addition, the fact that the suspension of MNPs is not insulated in applications makes this assumption a limitation of the adiabatic approach.

\noindent{\bf Approaching a system in non-adiabatic conditions.} 
Keeping in mind that the suspension of magnetic nanoparticles often interacts with the environment in applications and even in experiments, this fact cannot be neglected in a model addressing magnetic hyperthermia. 
Here, we propose a theoretical approach based on thermodynamics concepts that takes into account this interaction, thus improving the $SLP$ estimates. 
Specifically, we assume the interaction between system and environment is embedded in the contribution of the heat loss in the First Law of Thermodynamics, i.e.\ the $Q_{susp}$ term in Eq.~(\ref{1-lei}). 
Hence, we handle with a suspension of magnetic nanoparticles submitted to an alternating magnetic field in non-adiabatic conditions. 

\begin{figure*}[!ht]
\begin{center}
\includegraphics[width=12.cm]{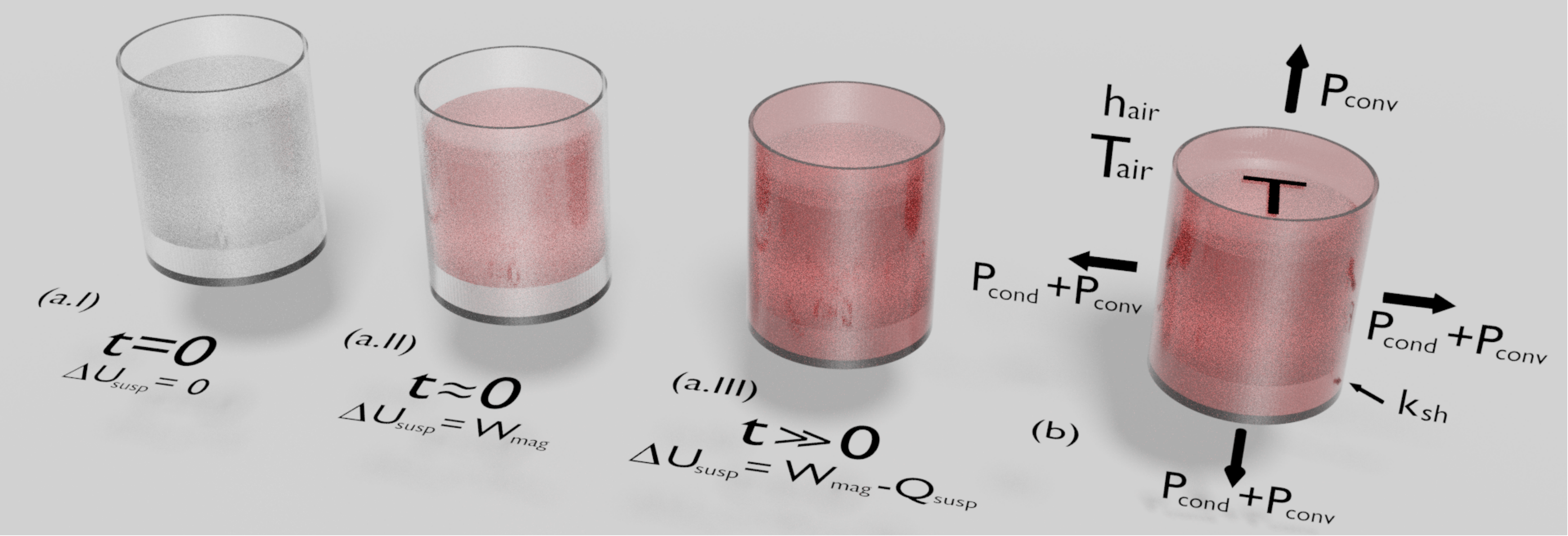}\\ 
%\vspace{-.1cm}\includegraphics[width=7.5cm]{Fig_01b.png}
\end{center}
\vspace{-.5cm}
\caption{Schematic representation of our theoretical system --- a suspension of magnetic nanoparticles inside a sample holder, which is submitted to an alternating magnetic field. Suspension (a.I) in thermal equilibrium with the environment, (a.II) in an adiabatic regime during a short time interval just after the AMF is turned on, and (a.III) in the non-adiabatic regime. 
(b)~Definitions of some quantities employed in our theoretical approach. Here, we consider $T$ as the temperature of the suspension, $P_{cond}$ as the heat loss rate due to the process of conduction through the walls of the sample holder, $\kappa_{sh}$ denotes the thermal conductivity of the sample holder, $P_{conv}$ corresponds to the heat loss rate to the convective process of heat transfer from the outer surface of the sample holder and the upper surface of the sample, both surrounded by air, $h_{air}$ is the heat transfer coefficient of the air, and $T_{air}$ is the temperature of the environment.}
\label{Fig_01}
\end{figure*}

Generally, our system consists of magnetic nanoparticles dispersed in a fluid, which is submitted to an alternating magnetic field. 
The suspension of magnetic nanoparticles is inside a sample holder, which plays as boundaries that split it from the environment, as we can see in Fig.~\ref{Fig_01}. 

In a MHT experiment, first, while the AMF is off, the system is in thermal equilibrium with the environment (Fig.~\ref{Fig_01}~(a.I)). 
As soon as the magnetic field is turned on, it acts on the system, and a magnetic work $W_{mag}$ is done on the suspension. 
In particular, at this stage, an adiabatic process is assumed; and this total irreversible work undergone by the suspension is converted to internal energy of the system, what is evidenced through an increase of the system temperature (Fig.~\ref{Fig_01}~(a.II)). 
We understand that the heat loss $Q_{susp}$ may be neglected during a quite-short time interval; 
 and, therefore, the approach for the system in the quasi-adiabatic regime become enough.
Hence, Eq.~(\ref{slp-adiab}) may be used carefully. 
However, after this interval in which the temperature varies linearly with time, the quasi-adiabatic approximation loses its validity. 
In this case, a fraction of the energy drawn from the magnetic field is converted into heat loss as well, giving rise to the energy exchange between suspension and environment (Fig.~\ref{Fig_01}~(a.III)).

Given all the stated above, we also start our approach from the the First Law of Thermodynamics, 
% given by Eq.~(\ref{1-lei}), 
\begin{eqnarray}
%	\Delta U_{susp} = W - Q_{susp},
	\Delta U_{susp} = W_{mag} + W_{mec} - Q_{susp}.
	\label{1-lei-completa_a}	
\end{eqnarray}

Here, although we assume a non-adiabatic regime, the process remains to be isochoric ($W_{mec}=0$), without thermal expansions and/or mechanical work done on the suspension. 
Then,
\begin{eqnarray}
%	\Delta U_{susp} = W - Q_{susp},
	\Delta U_{susp} = W_{mag} - Q_{susp},
	\label{1-lei-completa_b}	
\end{eqnarray}
\noindent where, just to remember, $\Delta U_{susp}$ is the variation of the internal energy of the suspension, $W_{mag}$ is the irreversible work {\it undergone by the suspension due to the interaction of the magnetic particles with the alternating magnetic field}, and $Q_{susp}$ is the heat {\it lost by the suspension}. 

From the differentiation of Eq.~(\ref{1-lei-completa_b}) with respect to time, we may express $SLP$ as 
\begin{eqnarray}
%	SLP =  \dfrac{1}{m_{np}} \dfrac{W_{mag}}{\Delta t}
	SLP  = \dfrac{1}{m_{np}} C_{susp} \frac{dT}{dt} + \frac{1}{m_{np}} \frac{dQ_{susp}}{dt}.
	\label{eq-dif}
\end{eqnarray}
\noindent Notice that the most suitable definition for the specific loss power in the generalized case, i.e.\ Eq.~(\ref{eq-dif}), is actually the total irreversible work rate per magnetic material mass undergone by the suspension.
%, given by Eq.~(\ref{slp-adiab}). 
%, minus the fraction due to heat loss rate per magnetic material mass, which is related to the energy leaving the system. 
As a consequence, Eqs.~(\ref{slp-adiab}) and (\ref{eq-dif}) are similar, except by the second term in the definition for the $SLP$ in the non-adiabatic regime. 
This latter denotes the dependence of $SLP$ with the rate of heat loss of the system to the environment, which we define here as $P \equiv \frac{dQ_{susp}}{dt}$. 

We first address here the heat loss rate due to the process of conduction through the walls of the sample holder. 
Then, taking into account the Fourier's Law~\cite{ChatoJohnCandPaulsenKeithDandRoemer2012}, it may be written as 
\begin{eqnarray}
P_{cond} = -\kappa_{sh} A_{sh} \dfrac{dT}{dr}, 
\label{eq-fourier2}
\end{eqnarray}
where $\kappa_{sh}$ and $A_{sh}$ are the thermal conductivity and surface area of the sample holder, respectively.
It is worth remarking that the heat loss rate is assumed to be normal to the surfaces of the system (See Fig.~\ref{Fig_01}(b)).
Additionally, for sake of simplicity, we use a convenient form of sample holder, with cylindrical form.
As a consequence, the variable $r$ denoting the radial distance, as well as $z$ expressing the height in cylindrical coordinates, changes in a direction normal to the system surfaces, and to the heat reservoir through the walls of the sample holder. 
This fact simplifies the solve of Eq.~(\ref{eq-fourier2}), without loss of generality. 
%and $ r $ (m) is a normal variable to the ``heat surface''. 
%is denotes the variable describing the position, i.e.\ the radial distance. 
%, and $r$ denotes axial distance, i.e.\ the position. 
%Notice in Fig.\ref{porta-amostras}d we assume that the heat loss rate is normal with respect to the surface of the sample, therefore, the position variable $ r $ (m) in Eq.\ref{eq-fourier2} grows in the normal direction from the sample surface to the heat reservoir through the wall of the sample holder.

%$T(R_{int}) = T$, $T(R_{ext}) = T_{ext}$, $T(z_{bottom,ext}) = T_{ext}$ and $T(z_{bottom,int}) = T$
Then, from Eq.~(\ref{eq-fourier2}), under the boundary conditions of $T(R_{int}) = T(z_{bottom,int}) = T$, $T(R_{ext}) = T_{ext}$ and $T(z_{bottom,ext}) = T_{ext}$, the heat loss rate due to the process of conduction through the walls of the sample holder may be expressed by 
\begin{eqnarray}
P_{cond} = \kappa_{sh} \left(\dfrac{A_{side}}{R_{ext} \ln(R_{ext}/R_{int})} + \dfrac{A_{bottom}}{L}\right) (T - T_{ext}), 
\label{eq-conducao}
\end{eqnarray}
where $A_{side}$ and $A_{bottom}$ are the lateral and bottom areas of the sample holder, respectively, $L= z_{bottom,ext} - z_{bottom,int}$ is the thickness of the wall, $R_{int}$ the inner radius, $R_{ext}$ the external radius, $T_{ext}$ is the temperature of the external surface of the sample holder, and $T$ is the temperature of the suspension. 

Next, we address the heat loss rate due to the convective process of heat transfer from the outer surface of the sample holder and the upper surface of the sample, both surrounded by air. 
In this case, by means of the Newton's Law of cooling~\cite{Bergman2011}, it can be expressed as
\begin{eqnarray}
P_{conv} = h_{air} A_{sh} (T_{ext} - T_{air}) + h_{air} A_{top} (T - T_{air}), 
\label{eq-convec}
\end{eqnarray}
where $ A_{top}$ is the upper surface area of the sample, $h_{air}$ is the heat transfer coefficient of the air, and $T_{air}$ is the temperature of the environment. 
%heat reservoir, portrayed by the environment. 
It is worth mentioning that the environment, i.e.\ the air, is assumed to have properties of heat reservoir, exhibiting $\frac{d T_{air}}{d t} = 0$. 

From Eqs.~(\ref{eq-conducao}) and (\ref{eq-convec}), we can define 
\begin{equation}
\epsilon_{sh} \equiv \kappa_{sh} \left(\dfrac{A_{side}}{R_{ext} \ln(R_{ext}/R_{int})} + \dfrac{A_{bottom}}{L}\right),
\end{equation}
\begin{equation}
\epsilon_{air,surf} \equiv h_{air} A_{sh}, 
\end{equation}
\noindent and
\begin{equation}
\epsilon_{air,top} \equiv h_{air} A_{top}. 
\label{conduct}
\end{equation}
\noindent Here, $\epsilon_{sh}$ represents the thermal conductance of the sample holder, $\epsilon_{air,surf}$ is the thermal conductance associated to the convection of the air on the external surface of the sample holder, and $\epsilon_{air,top}$ corresponds to thermal conductance associated to the convection of the air on the interface sample/air at the upper surface. 

As a result, after all the rate of heat loss of the system to the environment may be rewritten, in analogy with electric circuits, as 
\begin{equation}
	P = \dfrac{dQ_{susp}}{dt} = \epsilon (T - T_{air}),  
	\label{heat-loss}
\end{equation}
where 
\begin{equation}
\epsilon = \left(\frac{\epsilon_{sh} \epsilon_{air,surf}}{\epsilon_{sh} + \epsilon_{air,surf}}\right) + \epsilon_{air,top}
\label{epsiloneffective}
\end{equation}
is the effective thermal conductance into the surrounding of the sample. 
Notice that $T$ is the quantity probed in MHT experiments. 

At temperatures between $300$ and $320$~K, within the range required for biological applications, as well as for temperature right above this limit,  the heat loss rate due to the radiation is negligible~\cite{Teran2012}. 
Thereby, from Eqs.~(\ref{eq-dif}) and (\ref{heat-loss}), we obtain 
\begin{eqnarray}
	SLP = \dfrac{1}{m_{np}} C_{susp} \dfrac{dT}{dt} + \dfrac{1}{m_{np}} \epsilon (T-T_{air}),
	\label{eq-dif-sar}
\end{eqnarray}
\noindent which describes {\it the temperature of a suspension of magnetic nanoparticles submitted to an alternating magnetic field, taking into account the energy exchange between the system and the environment}. 
%Equation~\ref{eq-dif-sar}
The solution for the differential equation in the {\it heating process}, under the condition $T(0) = T_{air}$ depicting the suspension is initially at room temperature when the field is turned on, is
\begin{eqnarray}
	T(t) = T_{air} + m_{np} \dfrac{SLP}{\epsilon} \left(1-e^{-\frac{\epsilon }{C_{susp}}t}\right).
	\label{eq-T}
\end{eqnarray}
Notice that, at long time intervals, $t\rightarrow \infty$, the  suspension temperature reaches the maximum value 
\begin{eqnarray}
T_{max} = T_{air} + m_{np} \frac{SLP}{\epsilon}, 
\end{eqnarray}
\noindent corresponding to a steady state. 

In the case of the magnetic field is turned off after the heating, Eq.~(\ref{eq-dif-sar}) also provides the temperature response during the {\it cooling process}. 
To this end, assuming $SLP = 0$, the solution for the differential equation, under the condition $T(0) = T_{max}$ as the temperature of the suspension when the field is turned off, is
\begin{eqnarray}
T(t) = T_{air} + \Delta T_{max} e^{-\frac{\epsilon}{C_{susp}}t},
\label{eq-T-resf}
\end{eqnarray}
in which $\Delta T_{max} = T_{max} - T_{air} $ is the temperature difference between the suspension and the environment.
%is the difference between the temperature of the suspension when the field is turned off and the temperature of the environment. 
It is interesting to notice that $T_{max}$ in the cooling process is not necessarily the maximum temperature achieved in the steady state after heating, but it simply corresponds the initial temperature of the suspension anytime when the field is turned off. 

After all, given the stated above, the specific loss power can be directly measured from the experiments. 
As a straight consequence of Eq.~(\ref{eq-T}), $SLP$ may be simply expressed as 
\begin{eqnarray}
	SLP = \dfrac{\epsilon}{m_{np}} \dfrac{(T-T_{air})}{\left(1-e^{-\frac{\epsilon}{C_{susp}}t}\right)}. 
	\label{eq-slp}
\end{eqnarray}

It is interesting to verify that Eq.~(\ref{eq-slp}), by means of Taylor's expansion $ e^{- \frac{\epsilon}{C_{susp}}t} \cong 1 - \frac{\epsilon }{C_{susp}}t $, recovers Eq.~(\ref{slp-adiab}).
This feature reveals Eq.~(\ref{eq-slp}) is in fact a generalized form of Eq.~(\ref{slp-adiab}), both converging in the limit $ t\rightarrow 0 $, the quasi-adiabatic regime.
However, unlike Eq.~(\ref{slp-adiab}), the validity of Eq.~(\ref{eq-slp}) is not restricted to a short time interval after applying the field.
Therefore, our approach provides a feasible route to accurate estimates of $SLP$ for magnetic nanoparticles under AMF in non-adiabatic conditions. 

\noindent{\bf Comparison with the experiment. }
To confirm the validity of our theoretical approach, we analyzed the thermal behavior of magnetite and magnesium ferrite MNPs dispersed in water submitted to an AMF.
Our set of samples here includes superparamagnetic nanoparticles with distinct compositions and different particle sizes (see Methods section for details on the magnetic nanoparticles and experiments). 

%From this initial characterization, it is interesting to notice our set of samples includes superparamagnetic nanoparticles with distinct compositions, different particle sizes and, thereby, variable effective magnetocrystalline anisotropy. 
%From a initial characterization, it is interesting to notice our set of samples includes superparamagnetic nanoparticles with distinct compositions and different particle sizes. 
%From now on we focus on the thermal response of MNPs dispersed in water submitted to an AMF. 

In order to make easier a direct comparison between theory and experiment, as well as to verify the validity of our theoretical approach, we need to make use conventional units found in literature. 
To this end, we adopt the temperature in $\rm ^\circ C$, $m_{np}$ in $\rm g$, $C_{susp}$ in $\rm J/^\circ C$, $SLP$ in $\rm W/g$, $\epsilon$ in $\rm W/^\circ C$ and $t$ in $\rm s$. 
\begin{figure*}[!ht]
\begin{center}
\includegraphics[width=16.2cm]{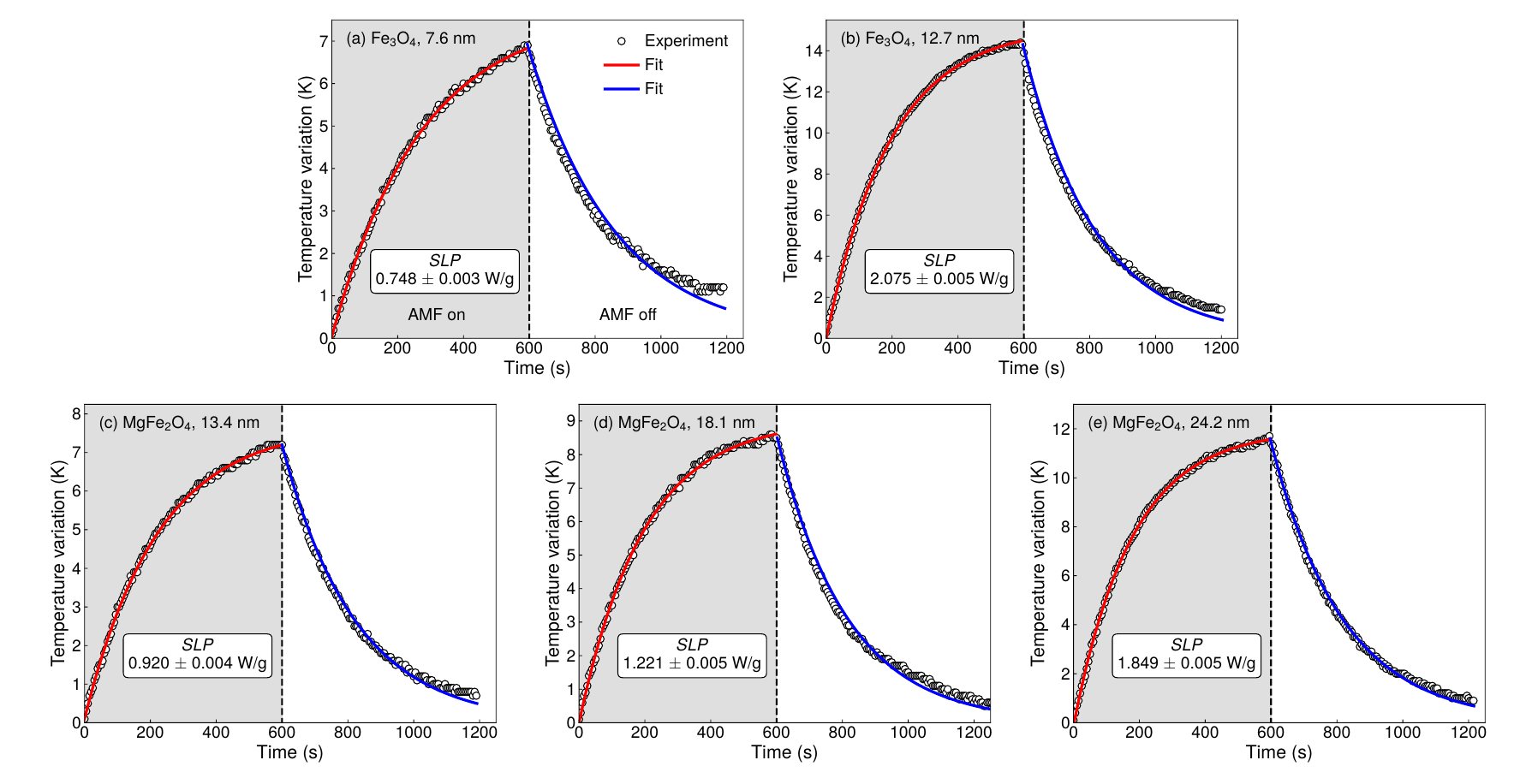} %18*0.9
\end{center}
\vspace{-.75cm}
\caption{Thermal response of our suspensions. Time evolution of the temperature of our magnetic (a,b) magnetite and (c-e) magnesium ferrite nanoparticles dispersed in water. The gray and white zones delimit the time periods corresponding to the heating and cooling processes, in which the suspension is exposed to an alternating magnetic field on and off, respectively. The magnetic hyperthermia experiments were performed with an AMF with frequency of $70.5$~kHz and amplitude of $70$~Oe. The symbols are the experimental data for the temperature as a function of the time. The red solid line is the data fit obtained using Eq.~(\ref{eq-T}), corresponding to the heating process. The blue solid line in turn is the fit for the cooling process, performed using Eq.~(\ref{eq-T-resf}), with $\Delta T_{max}$ being the maximum temperature variation achieved in the heating process. The values of $SLP$ and effective thermal conductance $\epsilon$ estimated from the fits are reported in Tab.~\ref{Tab_results}.}
\label{Fig_03}
\end{figure*}
\begin{table*}[!ht]\centering
	\caption{Summary of our findings. The average particle size were estimated by TEM. The experimental specific loss power and effective thermal conductance for our magnetite and magnesium ferrite nanoparticles were measured from the magnetic hyperthermia experiments performed with alternating magnetic field with frequency of $70.5$~kHz and amplitude of $70$~Oe. }%{\color{red}Sugiro inserirmos o desvio padrao para todas as quantidades.}}
	\begin{tabular}{ c c  c  c  c }
		\hline \hline
		\footnotesize{\textbf{Composition\hspace{.2cm}}} & \footnotesize{\textbf{\makecell{Particle \\ size (nm)}}} & \footnotesize{\textbf{\hspace{.2cm}\bm{$SLP$} (W/g)}} & \footnotesize{\makecell{\bm{$\epsilon$} \textbf{heating} \\ \textbf{(W/$^\circ$C)}}}  & \footnotesize{\makecell{\bm{$\epsilon$} \textbf{cooling} \\ \textbf{(W/$^\circ$C)}}} \\
		\hline \hline
		\footnotesize{\bf Fe$_3$O$_4$} & \footnotesize{$7.6\pm 0.2$} & \footnotesize{\hspace{.3cm}$0.748\pm 0.003$} &\footnotesize{\hspace{.3cm}$0.0100\pm 0.0001$\hspace{.3cm}} &\footnotesize{$0.0103\pm 0.0001$}  \\
		\footnotesize{\bf Fe$_3$O$_4$} & \footnotesize{$12.7\pm 0.2$} & \footnotesize{\hspace{.3cm}$2.075\pm 0.005$} &\footnotesize{$0.0138\pm 0.0001$}	&\footnotesize{$0.0123\pm 0.0001$}  \\
		\footnotesize{\bf{MgFe$_2$O$_4$}} & \footnotesize{$13.4\pm 0.3$} & \footnotesize{\hspace{.3cm}$0.920\pm 0.004 $} &\footnotesize{$0.0122\pm 0.0001$} &\footnotesize{$0.0122\pm 0.0001$} \\
		\footnotesize{\bf MgFe$_2$O$_4$}  & \footnotesize{$18.1\pm 0.2$} &\footnotesize{\hspace{.3cm}$1.221\pm 0.005$} &\footnotesize{$0.0136\pm 0.0001$} &\footnotesize{$0.0126\pm 0.0001$}   \\
		\footnotesize{\bf MgFe$_2$O$_4$}  & \footnotesize{$24.2\pm 0.2$} &\footnotesize{\hspace{.3cm}$1.849\pm 0.005$} &\footnotesize{$0.0153\pm 0.0001$} &\footnotesize{$0.0124\pm 0.0001$}  \\
		\hline \hline
	\end{tabular}
	\label{Tab_results}
\end{table*}

Figure~\ref{Fig_03} depicts the thermal response of our suspensions. 
Notice the quite-good concordance between experimental data and theoretical prediction. 
Given that $T_{air}$, $m_{np}$ and $C_{susp}$ are known experimental parameters, we first take into account the data from the heating process, and fit them using Eq.~(\ref{eq-T}). 
From this procedure, we estimate here the specific loss power and the effective thermal conductance into the surrounding of the sample. 
Next, we fit the data from the cooling process, thus considering Eq.~(\ref{eq-T-resf}) and assuming $\Delta T_{max}$ as the maximum temperature variation achieved in the heating process. 
From this latter case, we are able to confirm the effective thermal conductance obtained from the first fit procedure. 
Our findings are summarized in Tab.~\ref{Tab_results}. %, as aforementioned,

From the general point of view, all the main features of the time evolution of the temperature of magnetic nanoparticles dispersed in water submitted to an alternating magnetic field are well described by our approach to the magnetic hyperthermia in the non-adiabatic regime.
The tiny differences between experiment and theory, especially when the system is reaching the room temperature in the cooling process, may be devoted to small changes in the environment and/or modifications in the suspension due to the previous increase of the temperature, which are not taken into account in our model. 
Despite it, we obtain here consistent $SLP$ results. 
Specifically, we find values between $0.748$ and $2.075$~W/g for our suspensions, and we verify a clear raise of the specific loss power with the particle size, as expected. 

The dependence of the $SLP$ with intrinsic parameters of sample, such as average diameter, size distribution, morphology and crystalline structure of the particles~\cite{Rosensweig2002,Hergt2008,Usov2012} as well as viscosity of the fluid carrier~\cite{Zhang2007,Fortin2007}, has been previously verified by numerous groups.
Further, it is well-known the $SLP$ is dependent on the AMF, evolving in different form with frequency and amplitude. 
In this case, Hergt and colleagues~\cite{Hergt1998} have shown for aqueous suspensions of magnetite $SLP$ values between $0.1$ and $21$~W/g for a field with frequency of $300$~kHz and amplitude of $82$~Oe, while Zhang and coworkers~\cite{Zhang2007} have estimated for a similar system values within the range between $4.5$ and $75$~W/g for a field with $55$~kHz and $200$~Oe.
Therefore, our results are also in agreement with distinct findings reported literature. 

The most striking feature resides in the own $SLP$, as well as in its accuracy, i.e.\ the standard deviation of the values estimated with our approach to the magnetic hyperthermia in non-adiabatic conditions. 
From Tab.~\ref{Tab_results}, we may check the standard deviations of $SLP$ fall into the range between $0.003$ and $0.005$~W/g. 
To highlight our achievements, it is worth remarking that we also carried out an analysis of our experimental results considering the quasi-adiabatic method, thus employing Eq.~(\ref{slp-adiab}) to obtain the specific loss power. 
In this respect, we performed the fits considering the temperature variation of the suspension during the first $20$~s of the experiment. 
Besides obtaining underestimated $SLP$ results, between $\sim 0.60$ and $\sim 1.65$~W/g, the accuracy is substantially worse, with standard deviation values being one order of magnitude higher than those found from the non-adiabatic approach. 

Last but not least, we look at $\epsilon$, the parameter defined by Eq.~(\ref{epsiloneffective}) and named here as the effective thermal conductance into the surrounding of the sample. 
The $\epsilon$ values obtained from the fits of the experimental data in the heating and cooling processes using Eqs.~(\ref{eq-T}) and (\ref{eq-T-resf}), respectively, are shown in Tab.~\ref{Tab_results}. 
We understand the tiny variations in the values of the effective thermal conductance may be devoted to the fluctuations promoted by changes in the environment, modifications in the suspension due to the previous increase of the temperature, difference in surface area between the samples, as well as limitations of the own experimental setup. 
%We understand the tiny variations in the values may be devoted to fluctuations in the effective thermal conductance promoted by changes in humidity, temperature and air density, tiny differences in the surface areas of the samples, as well as limitations of the own experimental setup. 
%In this sense, we understand our experiments disclose similar $\epsilon$ values within the measurement error. 
In particular, we find the average $\epsilon$ parameter is $0.0125 \pm 0.0005$~W/g. 
It is worth mentioning that we have also performed tests considering distinct samples and fields having different amplitudes and frequencies, not addressed here, and all experiments uncover the very same $\epsilon$ parameter within the measurement error. 
Thereby, the small relative inaccuracy suggests this parameter is independent on the sample and/or magnetic field; but it is intrinsically related to the environment into the surrounding of the sample and the surface area of contact between the sample and environment, as expected whether Eq.~(\ref{epsiloneffective}) is indeed valid.
%{\color{red}Fazer algum comentario sobre medidas com diferentes razoes particula/agua, diferentes frequencias, etc? E colocamos algo sobre os tempos de relaxacao?}

After all, the quantitative agreement of preditions with experimental results do confirm the robustness of our theoretical approach. 
Hence, we provide physical meaning to parameters found in literature that still remained not fully understood, as well as bring to light how they can be obtained experimentally.
In addition, our findings place the theoretical approach to magnetic hyperthermia based on thermodynamics concepts, that takes into account the interaction of the system with the environment, as a sharp tool for the determination of an accurate, reliable specific loss power value from a non-adiabatic process. 

\section{Discussion}
\label{Discussion}

In summary, we have performed a theoretical and experimental investigation of the magnetic hyperthermia in suspensions of magnetic nanoparticles. 
Here we have proposed a theoretical approach to magnetic hyperthermia from a thermodynamic point of view. 
To test the robustness of the approach, we have performed hyperthermia experiments and analyze the thermal behavior of magnetite and magnesium ferrite magnetic nanoparticles dispersed in water submitted to an alternating magnetic field. 
By comparing experiment and theory, the model has allowed us to obtain the specific loss power of a suspension submitted to an alternating magnetic field from a non-adiabatic process. 
Remarkably, we have verified our approach enhances the accuracy in the determination of this quantity, when compared to quasi-adiabatic methods. 
%At last, 
We have also provided physical meaning to parameters found in literature that still remained not fully understood. 
Specifically, we have been able to address the effective thermal conductance, as well as the heat loss rate due to the conduction and convective processes, bringing to the light how they can be obtained experimentally. 
In this respect, regarding the effective thermal conductance, we have yet provided evidences that it is intrinsically related to the environment into the surrounding of the sample and the surface area of contact between the sample and environment. 
After all, it it worth remarking the quantitative agreement of preditions with experimental results has confirmed the validity of our theoretical approach. 
Thereby, our findings place the theoretical approach to magnetic hyperthermia based on thermodynamics concepts that takes into account the interaction of the system with the environment as a sharp tool for the determination of an accurate, reliable specific loss power value from a non-adiabatic process. 

\section{Methods}
\label{met}

\noindent{\bf Set of samples.} 
For the study, we prepared a set of $5$ samples. 
Two of them are magnetite $\rm Fe_{3}O_{4}$ nanoparticles, with particle size of $7.6$ and $12.7$~nm, synthesized by co-precipitation considering distinct proportions of precursor reagents~\cite{Witte2016,Lee2004}. 
The other three samples are magnesium ferrite $\rm MgFe_{2}O_{4}$ nanoparticles, produced by sol-gel followed by calcination at the selected temperatures of $400$, $500$ and $600^{\circ}$C for $2$~h~\cite{Hiratsuka1995,Benvenutti2009}. 
These latter have particle size of $13.4$, $18.1$, and $24.2$~nm, respectively. 
{\color{black}Thereby, our set is composed by nanoparticles having distinct compositions and different particle sizes.} 

\noindent{\bf Structural and morphological characterization.} 
The structural and morphological properties of the nanoparticles were verified by X-ray diffractometry (XRD) and transmission electron microscopy (TEM). 
The diffraction measurements were performed with a Rigaku MineFlex II diffractometer, and the results were refined by Rietveld method using the software MAUD, thus allowing the identification of the phase, and providing lattice parameters and crystallite size. 
TEM images were acquired with a JEM-1011 transmission electron microscope and analyzed using the software ImageJ, then informing the phase, particle shape and distribution of the average particle diameter. 
\begin{figure*}[!h]
%2x2 version
\begin{center}
\includegraphics[width=16.2cm]{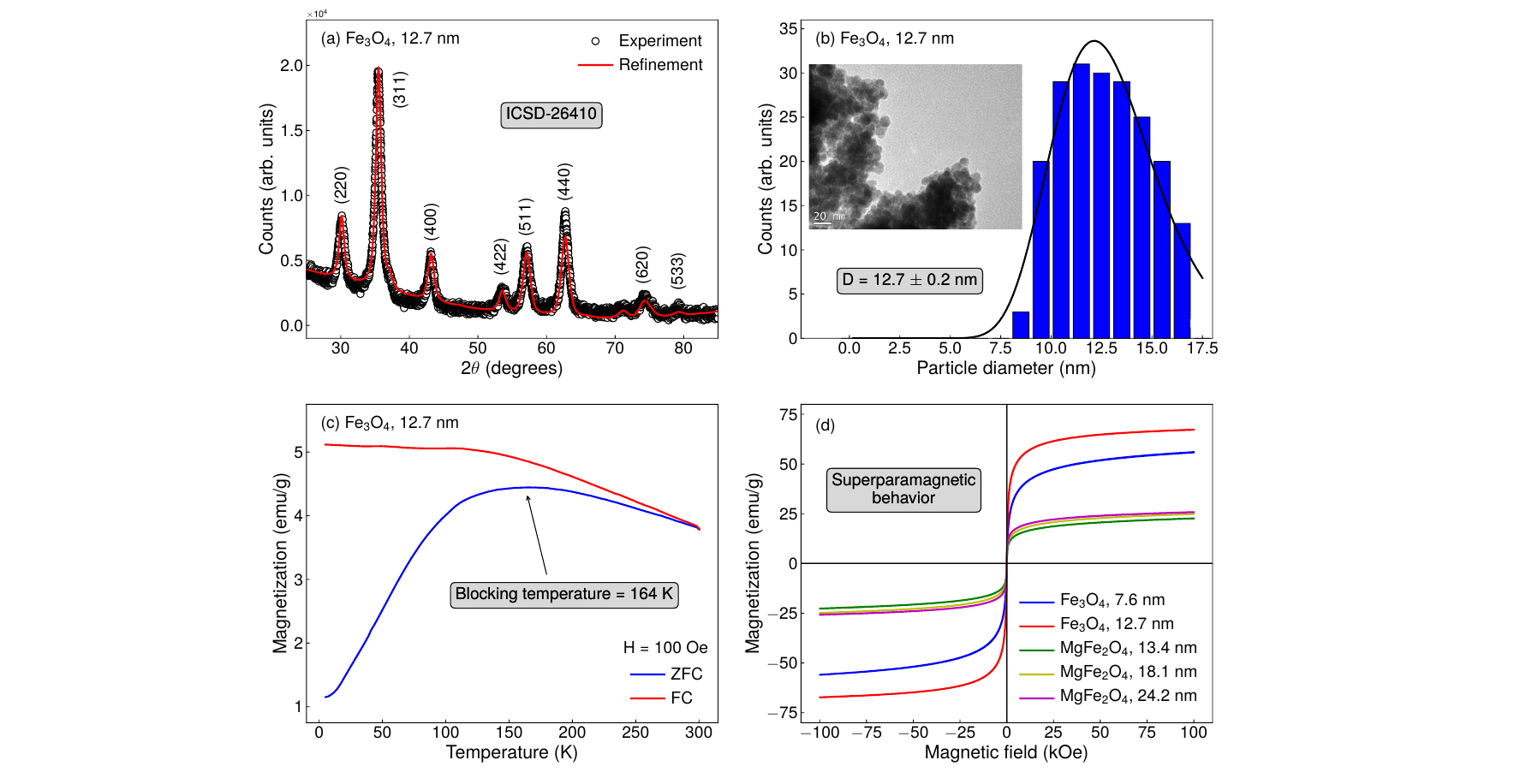} %18*0.9
\end{center}
\vspace{-.75cm}
\caption{Structural, morphological and magnetic properties of our magnetite and magnesium ferrite nanoparticles. (a) High angle X-ray diffraction pattern with Rietveld refinement, (b) transmission electron microscopy image with histogram of particle size distribution fitted with a log-normal function, and (c) ZFC and FC magnetization curves acquired with probe magnetic field of $100$~Oe for the magnetite sample with average particle diameter of $12.7$~nm, as representative examples of our findings for the investigated nanoparticles. (d) Isothermal magnetization curves measured at room temperature for the magnetite and magnesium ferrite samples with distinct particle sizes.}
\label{Fig_02}
%1x3 version
%\begin{center}
%\includegraphics[width=18cm]{Fig_02.pdf} 
%\end{center}
%\vspace{-.75cm}
%\caption{Structural, morphological and magnetic properties of the magnetite and magnesium ferrite nanoparticles. (a) Transmission electron microscopy image with histogram of particle size distribution and (b) ZFC and FC magnetization curves acquired with probe magnetic field of $100$~Oe for the magnetite sample with average particle diameter of $12.7$~nm, as representative examples of our findings for the investigated nanoparticles. (c) Isothermal magnetization curves measured for the magnetite and magnesium ferrite samples with distinct particle sizes.}
%\label{Fig_02}
\end{figure*}

Figures~\ref{Fig_02}(a,b) bring representative examples of the results obtained from the structural and morphological characterization. 
From the XRD experiments, we first confirm our samples are single phase. 
%Specifically, diffraction peaks for our magnetite samples are well indexed with the reference pattern ICSD-26410 of the Fe3O4 with cubic symmetry and Fd:3m space group.
%The peaks for our magnesium ferrite samples in turn, not shown here, are in quite-good concordance with the pattern ICSD-152468, also revealing cubic symmetry with Fd:3m space group. 
Specifically, diffraction peaks for the magnetite samples are well indexed with the standard pattern ICSD-26410, and can be associated to the ($220$), ($311$), ($400$), ($422$), ($511$), ($440$), ($620$), ($533$) planes. 
These findings are in very good agreement with reports found in the literature~\cite{PB472p66, CSA560p376, JDST0p1}. 
The results for the magnesium ferrite in turn are in quite-good concordance with ICSD-152468 and with findings previously reported by different groups~\cite{JMMM320p2774,JEN8p347, APA124p319,CJP56p2218}, presenting peaks located at $2\theta$ ranging from $28^\circ$ to $80^\circ$, which are associated with the ($220$), ($311$), ($222$), ($400$), ($422$), ($511$), ($440$), ($620$) and ($533$) planes of the MgFe$_{2}$O$_{4}$. 
For both compositions, the patterns raise fingerprints of phases having cubic symmetry and Fd:3m space group. 
{\color{black}Rietveld refinement yet informs us the crystallite size, confirming our procedures as promising routes to the production of pure nanoparticles with specific sizes.} 
All these findings are corroborated by TEM. 
TEM images also show the particles are quite uniform, having approximately spherical geometry, despite the clusters formation. 
The histograms of particle size distribution fitted with a log-normal function 
%reveal small particle size dispersion for all samples, and 
confirm the aforementioned average particle diameter values between $7$ and $25$~nm. 
Table~\ref{Tab_results} discloses specifically our findings on the particle size for each sample. 

\noindent{\bf Magnetic characterization.} 
The magnetic characterization of the nanoparticles were performed using a Quantum Design Dynacool Physical Property Measurement System through zero-field-cooled (ZFC) and field-cooled (FC) magnetization measurements, acquired in the range of temperature between $4$ and $300$~K with probe magnetic field of $100$~Oe, as well as via isothermal magnetization curves acquired at selected temperatures.

Figure~\ref{Fig_02}(c) shows a representative example of the ZFC and FC magnetization curves measured for our samples. 
%All the main features of both curves delineating the magnetization as a function of temperature are well understood. 
All the main features of both curves representing the dependence of the magnetization with temperature are well understood. 
From our concern at this moment, we highlight the ZFC curves are characterized by a broad cusp, whose location of the maximum defined the system blocking temperature, in which the nanoparticles exhibit a magnetic transition between the superparamagnetic and blocked states. 
For our set, we find blocking temperature values within the range between $135$ and $190$~K. 
Similar results are found in literature for both, magnetite~\cite{PB472p66,CSA560p376} and magnesium ferrite~\cite{JMMM194p1, JMS30p768, JAC699p521, CJP56p2218, AIP1665p050095} nanoparticles. 
{\color{black}In this sense, our samples are superparamagnetic at room temperature.} 
Figure~\ref{Fig_02}(d) presents the magnetization curves acquired for our magnetite and magnesium ferrite samples with distinct particle sizes. 
Remarkably, all samples exhibit a typical behavior of a soft magnetic material. 
Below the blocking temperature, the isothermal magnetization curves, not shown here, exhibit hysteresis, as expected. 
At room temperature, we observe s-shaped curves, with low remanent magnetization and small values of coercive field, being well described by a Langevin function, characterizing the superparamagnetic state. 

\noindent{\bf Magnetic hyperthermia experiments.} 
The calorimetric measurements were carried out with a homemade experimental setup. 
The system consists basically of two parts, one responsible by generating of the AMF and another by the detection of the sample temperature. 
The first one is composed by a parallel LC resonant circuit~\cite{Boylestad1966}, which includes the solenoid and provides a homogeneous sinusoidal magnetic field with frequency of $70.5$~kHz and amplitude of $70$~Oe. 
We took special care to minimize effects due to Joule losses during the measurements. 
In this respect, a cooling system is responsible by keeping the solenoid at room temperature. 
The second part of the system consists in an Extech HD300 infrared thermometer, which allows us to perform precise acquisitions of the sample temperature. 
All the measurements were performed in suspension samples, consisting of $100$~mg of nanoparticles dispersed in $0.6$~mL of distilled water. 
Specifically, we divided the experiment in two stages. 
In the first stage, once the suspension was at room temperature, we turned on the AMF and started acquiring the sample temperature. 
After recording the temperature in the heating process during $600$~s, the second stage begins when the field is turned off, and we kept the temperature measure for an additional period of $600$~s during the cooling process.

\section*{Acknowledgments}
~The research is supported by the Brazilian agencies CNPq, CAPES, and FAPERN. 
The authors would like to thank the LCME-UFSC for the technical support during the electron microscopy procedures (LCMEMAT/2020). 
L.~Streck and F.~Bohn dedicate this work especially to Gabriela Streck Bohn.

\section*{Author Contributions}
~C.A.M.I., E.L.B., L.S., J.L.C.F., F.B.\ prepared the set of samples. 
C.A.M.I., J.C.R.A., J.X., R.B.S., J.M.S., C.C.P.C., M.G., E.F.S., C.C., M.A.C.,S.N.M., F.B.\ performed the experiments
C.A.M.I., and F.B.\ were responsible for the theoretical approach. 
C.A.M.I., J.C.R.A., J.X., R.B.S., and F.B.\ interpreted the results and wrote the original text of the manuscript. 
All authors contributed to improve the text.

\section*{Competing Financial Interests statement}
%~The authors declare no competing financial interests.
~The authors declare no competing interests.  

\section*{Additional information}
~Correspondence and requests for materials shall be addressed to F.B.\ and C.A.M.I.

\end{document}